# Query Optimization Over Web Services Using A Mixed Approach

Debajyoti Mukhopadhyay[1], Dhaval Chandarana[1], Rutvi Dave[1],
Sharyu Page[1], Shikha Gupta[1]

[1]Maharashtra Institute of Technology, Pune 411038
India
{debajyoti.mukhopadhyay, chandaranadhaval, rutvi.dave12, sharyu69,
shikhagupta.172}@gmail.com

**Abstract.** A Web Service Management System (WSMS) can be well-thought-out as a consistent and a secure way of managing the web services. Web Service has become a quintessential part of the web world, managing and sharing the resources of the business it is associated with. In this paper, we focus on the query optimization aspect of handling the "natural language" query, queried to the WSMS. The map-select-composite operations are piloted to select specific web services. The main aftermath of our research is ensued in an algorithm which uses cost-based as well as heuristic based approach for query optimization. Query plan is formed after cost-based evaluation and using Greedy algorithm. The heuristic based approach further optimizes the evaluation plan. This scheme not only guarantees an optimal solution, which has a minimum diversion from the ideal solution, but also saves time which is otherwise utilized in generating various query plans using many mathematical models and then evaluating each one.

## 1 Introduction

Consider a scenario; a client fires a complex query on an SQL like interface, over the internet. In order to retort the query, the query processor needs to access various manageable resources and interfaces of the relatable business organizations. Here the web service comes into picture.

A web service is chosen for managing the resources of a company or a business organization over the internet. This is because it is:
1) Language and platform independent
2) Allows encapsulation
3) Has wire level standards
4) Follows industry momentum.

Therefore a web service enables application to application interaction over the web irrespective of their platform. It is provided in the middle tier of application servers. The database of a company is put behind a web service in an organized fashion.

Compared to database, web services are fast growing, more heterogeneous and change more rapidly. Also while querying web service semantic information, like the data type of the output isn't enough. Specific semantics need to be mentioned. Also multiple web services can be queried based on their dependency constraints.

Web Service Management manages the web services in an organized and secure manner.

Web Service Registration allows registration of new web services. It uses UDDI (Universal Description, Discovery and Integration) which manages web services accessed by the mass individuals and provides the registry with WSDL (Web Service Description Language) for the same. The UBR (Universal Business Registry) manages these Web Services using taxonomy.

A web service provider registers a web service on UDDI. Based on the web service type (Definition) and the service implementation (instance) the UDDI stores the web service using a data model. The web service requestor or the consumer can call a web service if he knows what to search for on the UDDI.

The WSMS acts like a broker between the web service requestor and web service provider.

It has an SQL like interface which allows the user to shoot a complex, multi-query at a time. The query processor in a WSMS proposes a possible evaluation plan and the optimizer attempts to determine which plan will be the most efficient.

The WSMS acts like a broker between the web service requestor and web service provider.

It has an SQL like interface which allows the user to shoot a complex, multi-query at a time. The query processor in a WSMS proposes a possible evaluation plan and the optimizer attempts to determine which plan will be the most efficient.

The efficiency of a query evaluation plan can be determined by the cost of executing the same. It can be measured in terms of different resources including disk accesses, CPU time to execute a query and in a distributed or parallel system the cost of communication.

The response time for a query evaluation plan i.e. the time required to execute the plan, assuming no other activity is going on. Profiling is based on the response time, characteristics and statistics of a web service call. The information is stored by WSMS and used while estimating the cost of execution.

In this paper our main focus is on the query optimizer component of the WSMS. It treats queries as function calls and focuses on the performance not QoS.

A user fires queries on an interface. This query is expressed in a high level query language. This query needs to be scanned, parsed and validated. The scanner breaks down the query into tokens for instance SQL keywords, attribute names and relation name. The parser checks for any grammatical error in the query and validation of the attribute and relation name is done.

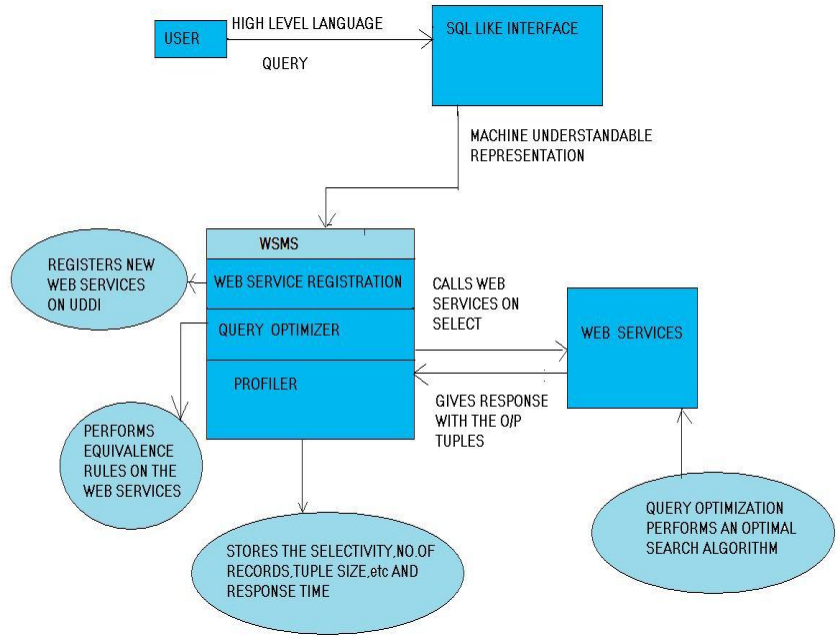

**Figure 1 System architecture.**

In WSMS the resources are managed by web services, as mentioned earlier. A web service has a number of operations called the service instances. In addition to this there can be a number of web services having the same functionality. The tokens obtained by breaking the query are sent to the web services which execute them according to their operations and return the results.

One web service may or may not satisfy the functional need for an activity. Thus the integration of many web services is done in order to satisfy the requirements. While doing so the precedence constraints need to be fulfilled. The precedence constraint controls the order in which each web service is executed.
 Relational-algebra operations are used to evaluate the query and are called an evaluation primitive. A sequence of these evaluation primitives creates the query evaluation plan.
After the scanning, parsing and validation, the query evaluation can be represented as a relational-algebra or a parse tree. The choice of the annotation completely depends on the type of database used.
We assume that our database represents the query evaluation plan using a parse tree.

Query optimization is the process of selecting the most efficient query evaluation plan among the many strategies usually possible for processing a given query.

There have been several attempts achieve query optimization over web services. Our methodology in achieving a solution was to first analyze the limitations of the same and then overcome them while designing a new strategy.

The following section (Section 2) will define and explicate our proposed model and algorithm.

Section 3 contains the related work on this research.

Section 4 will state its advantages over other methods.

## 2    Model

It is the responsibility of the system to construct a query evaluation plan that minimizes the cost of query optimization. Not only the query should be answered correctly, the system should make sure that the query plan is executed in a given time frame. This makes query optimization one of the most fundamental parts of any WSMS. Web service performs many operations. Each of these operations can be considered as an instance of a web service. In addition to this more than one web service can have the same functionality. A web service schema can be represented by a service graph with the instances as the node and dependency between them shown by the edges. As in a relational database management system the algebraic operator correspond to the low level primitives of a physical system like files, records and tuples the query algebra in WSMS depends on the shape of the service graph and the instances of the web service. In order to select the web service and form an evaluation plan, following algebraic operations can be considered:

Map: Mapping the Web service depending on the services required by the user.

Select: Calculate the response time of each web service and select the web service depending on the minimum response time.

Composition: When more than one web service is used to answer the user query, composition of web service is formed.

After selecting the most optimal web service to respond to a specific query, the evaluation plan is formed. The service execution path arranges all operations in the service graph into a sequence with respect to all the dependency or precedence constraints. Precedence Constraints can be explained as follows: A Web Service (WS2) only takes a particular tuple as input, and if that tuple is only given as output by another Web Service (WS1), then we can say that a precedence constraint exists between WS1 and WS2 OR WS1 always precedes WS2 in any Query plan, i.e. WS1 ☐ WS2.

Each of the operations could belong to one or different web service, and an executable service instance, which is provided by the web service provider, needs to be called in order to implement the plan. After the operations and their dependency constraints have been taken into account, the following steps take place:

1.    Apply the selection predicate to each of the relevant web service operation.

2.  After retrieving the data from the web services store the output as database on a local server and perform low level relational algebraic operations on them, 'select project and join'.
3.  Join operation is performed to combine and filter the data obtained from the various service operators.
4.  Project is used to display the final output to the user through the interface.
5.  Equivalence rules are applied to further optimize the plan.

The user need not know about the way web services are accessed till he receives the desired output. WSMS not only integrates use of more than one web services to solve a user query but also optimizes its cost based evaluation plan.

## 2.1  Algorithm

1.  Accept the query of the user through an SQL-like Interface.
2.  Use the service query algebraic operators to select one or more web services.
    - Map is used to find the web services that satisfy the required functional need
    - Select is used to find cost efficient web service among the mapped web services using a greedy method.
    - Composite combines many web services required to answer the query.
3.  Form the query evaluation plan using the dependency constraints among the service instances.
4.  Apply predicate to the web service
5.  Call the web services
6.  Store the retrieved data in primitive physical storage at local server.
7.  The Parse tree representation of the fired Query is generated using the data and low level relational operations.
8.  Relational algebraic operations are performed
    - Use greedy approach to perform join operation.
    - Use greedy approach to perform select operation
9.  Perform Optimization of sub trees (having two web services and one join) using Equivalence Rules and Report to WSMS
10. If the Parse Tree is completely evaluated then
    - PROJECT the Output tuples.
    - Report to WSMS.
    Else go to 8.

## 2.2  Selection Strategies

Suppose the web service management system wants to call a web service. This call consumes resources at both client and the server side. The client will initiate the service call sc. This call involves the following steps:

1. Initializing WSMS (the client) i.e. Ci modules that are responsible for activating sc.
2. Packing sc's parameters into the web service input message
3. Sending the call message to the server having the web service
4. Waiting for the server to process its request and send the result back; and
5. Unpacking, parsing, and inserting the result in the client.

Therefore the respective cost components of the service call:

cost (sc) = initiate+ callsize * (packing) + (callsize+packetsize) * (sending) + scost + resultsize * (unpacking).

The server side processing includes both the costs of service provider processing and network transfers to return the result to the client. On server side, the total cost for handling the service call can be given as:

scost (sc) = initiate+ callsize * (unpacking) + serviceexec + resultsize * packing + (resultsize + packetsize) * sending.

**Table 1 Service Call Primitives**

| Initiate | Time to initialize software modules at, respectively, service client and provider |
|---|---|
| callsize and resultsize | Size (in bytes) of, respectively, the service call and its result |
| packing and unpacking | Average time of, respectively, SOAP packing and unpacking |
| Packetsize | Average size (in bytes) of the SOAP packet |
| sending | Average time to transfer bytes from client to server |
| scost | Time spent between the SOAP message containing sc is received by the service provider, and sc result is received at the client |
| Serviceexec | Time of the service call execution on the provider |

**2.3   Heuristics based Optimization**

Cost based optimization is very helpful in finding out the most cost efficient evaluation plan. But the drawback in this method is the cost of optimization itself. The number of evaluation plans generated is so huge that finding the optimal plan itself requires a lot computational effort. These drawbacks can be removed by using the heuristics to reduce the cost of optimization. An example of a heuristic is that:
• Perform selection as early as possible

Basically performing the selection before any operation reduces the size of the relation by selecting only those tuples which are required in the next operating and discarding the others. Although these heuristics does not always reduce the cost but it helps to make the processing of further operations easier.

Similarly projection operation also reduces the size of the relations. So whenever we need to generate a temporary relation, we can use projections earlier. This gives another heuristic:
- Perform projections earlier.

### 2.4 Equivalence Rules

There can be an enormous difference in the cost of the evaluation plan. One of the factors is the selectivity of the joins which are used to combine the web services. We can generate logically equivalent expressions using equivalence rules. Annotations of these resultant expressions will give us different plans. Now the cheapest plan can be selected from these plans according to the estimated cost.

Equivalence rules say that the expressions of two forms are equivalent. These two expressions produce the same multiset of tuples. So the basic idea is to use these equivalence rules to find the alternate plans which are most cost efficient.

There are around 12 equivalence rules that are possible. If we consider all those rules and apply them to our query, there will be a huge number of plans that will be possible. Handling that huge number is practically impossible.

The main thing about our algorithm is that we are using the bottom up approach in which we are starting our optimization from the least subtask. This means we are optimizing every single subtask possible. So at a time we are always dealing with only two web services which implies that at a particular level we will use an equivalence rule that considers only two web services. Thus all the other equivalence rules can be ignored. The greedy approach will now have to work with a very less number of rules which will reduce out execution time to a great extent.

## 3 RELATED WORK

Web service management system is a vast topic. Our scope is limited to the area of optimization done over the web services. A lot of research has already been done in this area.

In [1], a Web Service Management System (WSMS) is proposed to enable optimized querying of Web services. An algorithm is proposed to optimized access Web services where a fixed input which is a classical Select-Project-Join query over Web services in order to retrieve a variable output. It arranges Web services in a query based on a cost model and returns a pipelined execution plan with minimum total running time of the query. In our web service optimization model, we use the SPJ to form the evaluation plan as in [1] but the optimization at relational as well as query level algebraic operation is done using a greedy approach.

The concept of precedence constraints used in [2] is also used in this paper. A Web Service (WS2) only takes a particular tuple as input, and if that tuple is only given as output by another Web Service (WS1), then we can say that a precedence constraint exists between WS1 and WS2 OR WS1 always precedes WS2 in any Query plan, i.e. WS1 -> WS2.

A WSPRC model mentioned in [3] is used to for building an efficient Optimizer. It is short for - Web Service Profiler Re-optimizer Cache. This is an adaptive model for query optimization over web services. The core component of the model is the re-optimizer which is primarily based upon adaptive greedy algorithm. Adaptive greedy improves the efficiency of querying optimization over web services.

Adaptive query processing is basically adjusting the query process dynamically based on the query feedback. This process is successful to optimize over multiple queries and distribution of data behind web services.

The main idea underlying the Web Service Profiler-Reoptimizer-Cache model is that when users try to find Web services through the model, it will execute a query for retrieving Web services and analyze them. Meanwhile, the model optimizes the query process by developing a query plan and saving it in the cache. Web services will be invoked in accordance with the query plan.

Adaptive Query Processing is based on an adaptive loop framework which includes four components: measurement analysis, planning and actuation.

1. Constructing the initial web service query plan: This plan is basically according to the sequence of invokes. It is mostly a parallel plan.
2. Classification of web services: Evaluation and analysis of web service query cost is done.
Various factors are considered:
   a. Root level and precedence constraints.
   b. Without precedence constraints but with relevant predicates.
   c. Without precedence constraints and without relevant predicates.

This step is repeated until all the required web services are arranged in some particular optimized manner.

This query plan is then executed and the results are sent to the user. At the same time this query plan is also saved in the result cache for future reference.

Now as we know that the web services and the data information are always undergoing some changes so there is a need of re-optimization.

3. Re-optimization caused by input information change or web service change: The measurement component of adaptive loop framework will continuously check the input tuples and the web services for some changes. If any of them cause an increase in the cost, it would adjust the web service query plan using adaptive greedy algorithm. This plan is now again saved in the cache.

We have only combined the basic Greedy algorithm with the heuristic techniques. But future work on combining Heuristics and Equivalence Rules with other optimizing algorithms could be done.

We have used the basic Greedy algorithm to find the best technique to solve the algebraic operations instead of Adaptive greedy. The Greedy approach is a highly suitable option for any WSMS and thus it has combined with some Heuristics for further optimization.

[4] Proposes an Automatic Web services composition that has attracted much attention in recent. Here more than one web service can be used in an integrated manner to answer the user query. This concept is used in the model created in this paper.

A query optimizer focuses on the performance of the web service than quality of web service. [5] Offers that interoperability and flexibility, the performance is a key research point of the Web Services application.

The service query framework presented in [6] uses a query algebra, whose concept has been referred while implementing our model. In [6] through service-oriented queries a user could access the web services.. The algebra defines a set
of algebraic operators also used in our paper. Algebraic service queries can be expressed using these operators allowing the users to query their desired services based on both functionality and quality. Various SEPs generated are based on dynamic programming and greedy algorithm, whereas in the mixed approach explained in this paper a single SEP is formed.
.
The selection strategy, search space and use of some heuristics for optimization mentioned in [7] are referenced in the paper.

## 4   ADVANTAGES

1.   This algorithm spends too much time in deciding an evaluation plan which is to be passed to the execution engine.  But the difference in cost (in terms of evaluation time) between a good strategy and a bad strategy is often substantial, and may be of several orders of magnitude. Hence it is worthwhile for the system to spend a substantial amount of time on the selection of a good strategy for processing a query, even if the query is executed only once.
2.   The other biggest advantage of this strategy is that it finds the most optimal solution and sub solutions which can be used while evaluating further queries.
3.   The strategy is devised such that the plan changes according to the input tuple, thus looking at every feature before evaluating the plan.
4.   A single query plan is formed. This saves time which is usually spent in building many queries and evaluating them

5. This plan also reduces the work by discarding the complex and unnecessary equivalence rules. It uses only those rules which include only one or two operands.

# 5 CONCLUSION

We have proposed a basic WSMS model, which takes a query in SQL form and provides the most optimal solution. The basic goal of this system is that the client should receive the most relevant solution to his query, within least amount of time. This is achieved through the cost based approach to select the best web services and the Greedy Algorithm combined with a set of Heuristics and Equivalence Rules. This is very simple and highly efficient. We use Greedy method at individual node-level of the tree, which simplifies query processing and also facilitates the use of Equivalence Rules as only two web services will be performing an operation at a particular time. This combined with very basic Heuristics, such as "Select operation should be performed as early as possible", enhances the performance significantly. This algorithm should very easily be implemented in the existing Web Service Management Systems.

Future Work related to this paper could be:1)We could replace the basic Greedy Algorithm with a more complex algorithm to decrease the cost further. 2)Work could be done on using our model for querying Web Services as per their QoS. 3)Deeper study about performance evaluation and the factors affecting it can be done.